# ROSAT/PSPC observation of the distant cluster CL0939+472

**Sabine Schindler**[1,2] **and Joachim Wambsganss**[3]

[1] MPI für extraterrestrische Physik, Giessenbachstr., 85748 Garching, Germany; e-mail: `sas@mpa-garching.mpg.de`
[2] MPI für Astrophysik, Karl-Schwarzschild-Str. 1, 85748 Garching, Germany
[3] Astrophysikalisches Institut Potsdam, An der Sternwarte 16, 14482 Potsdam, Germany; e-mail: `jwambsganss@aip.de`

**Abstract.** We present results and an analysis of a ROSAT/PSPC pointed observation of the galaxy cluster CL0939+472. This very rich and very distant cluster is known to contain an unusually high fraction of blue and E+A galaxies, and a lot of merging of galaxies is going on. The X-ray properties we find are also quite unusual. The luminosity in the ROSAT band (0.1-2.4 keV) is $L_X = (6.7 \pm 0.3) \times 10^{44}$ erg/s, rather on the low side for such a rich cluster. We find a temperature of $2.9^{+1.3}_{-0.8}$ keV, which is also relatively low, but has also a large uncertainty. The cluster shows indications of substructure, which results in a high value for the core radius and a steep drop off outside the core, if one fits a spherically symmetric $\beta$-model. The lower limit on the mass inside a radius of $1.0 h_{50}^{-1}$ Mpc is $2.6 \times 10^{14} M_\odot$, and the upper limit on the gas mass fraction is 50%. All the unusual properties of this cluster can be explained if in fact we see the merging of two clusters of roughly equal mass, which is also supported by recent optical observations.

## 1. Introduction

The distant galaxy cluster CL0939+472 (also known as Abell 851) is extremely rich, probably the richest cluster known (Dressler et al. 1994a). At a redshift of z = 0.41, it is the most distant Abell cluster. At the same time it is one of the optically best studied galaxy clusters at high redshift, by now about 30 redshifts are measured spectroscopically, and about 170 galaxies are assigned morphological types and template fitted redshifts from low resolution spectra (Belloni & Röser 1996). The cluster shows various signs of activity; a large fraction of the galaxies are blue galaxies, and a lot of merging and interaction is going on. Furthermore, a high redshift quasar is seen close to the centre of the cluster, apparently surrounded by a subgrouping of small blue galaxies. All this makes CL0939+472 a prime target for X-ray study. The fact that in optical light many member galaxies of high redshift clusters look more similar to nearby field galaxies than to nearby cluster galaxies (Dressler et al. 1994a) makes it even more interesting to study the X-ray properties of this cluster.

X-ray observations of galaxy clusters provide especially good information on the structure of clusters and thereby on their dynamical state. The X-ray surface brightness is particularly sensitive to changes in the potential since the X-ray emissivity varies as the square of the gas density. The PSPC on ROSAT (Trümper 1983) is well suited for the study of clusters because of its high sensitivity and its high spatial resolution with simultaneous energy resolution. These properties allow to estimate e.g. the dark matter distribution within clusters and the baryon to dark matter ratio.

Here we present an analysis of a ROSAT/PSPC observation of CL0939+472. We show the morphology of the X-ray emission, determine the X-ray profile, give estimates of the mass of the X-ray emitting gas and the total mass in the cluster. We discuss our results and compare this cluster with other X-ray clusters.

## 2. The cluster CL0939+472 (Abell 851) in optical light

The cluster CL0939+472 is among the seven clusters that are extensively studied by Dressler & Gunn (1992). They present photometric data for about 300 cluster galaxies, and determine the mean redshift and the velocity dispersion from 31 spectra to z = 0.4069 and $\sigma$ = 864 km/sec, respectively. They emphasize the extraordinary richness of this cluster, with roughly 500 galaxies brighter than $M_r \approx -18$.

Subsequently, Dressler et al. (1993) present results of a 6h observation with the original Wide Field/Planetary Camera of the Hubble Space Telescope (HST WFPC1). They point out a group of about 30 very small blue objects with magnitudes $22 < r < 25$ which apparently concentrate around a redshift z = 2.055 quasar. These galaxies are strikingly different from other cluster galaxies: they are very blue and of subarcsecond size (though clearly non-stellar). Their extent corresponds to physical sizes of few kiloparsecs. The a posteriori probability for this group to be a chance association was determined to be

*Send offprint requests to*: Sabine Schindler

$3 \times 10^{-4}$. Dressler et al. (1993) suggest that this association could be a cluster of nascent galaxies at high redshift, possibly physically associated with the quasar.

In a further study of CL0939+472 Dressler et al. (1994a) analyse the morphology of this cluster, based again on WFPC1 observations. Assigning Hubble types to galaxies brighter than $r \approx 24.0$, they find a significant fraction undergoing tidal interactions or mergers: as a conservative estimate they classify 24 out of 135 galaxies brighter than r = 23.0 to be probable or possible merger or interaction products. They suggest that interactions between galaxies could cause the strong signs of starbursts seen in some blue galaxies, when they result in mergers. However, they furthermore conclude that most of the excess blue galaxies in this distant cluster are late-type spirals.

In one of the first deep HST observations with the repaired Wide-Field/Planetary Camera (WFPC2), Dressler et al. (1994b) took a very deep high resolution image of this cluster. They are able to classify as many as 181 galaxies, of which 23 show indications of interactions with nearby galaxies. Comparison of these new high quality morphological types with those obtained with the older WFPC1 ones shows that there are no systematic differences.

The earlier discovery that the Butcher-Oemler effect (Butcher & Oemler 1978) is the consequence of a very high fraction of late-type spirals and irregulars is confirmed with the WFPC2 observation. Dressler et al. (1994b) also see a lot of evidence for mergers and interactions in CL0939+472; not enough, however, to explain the rareness of the spirals today. They suggest disk destabilization due to the spirals' halos being tidally distorted or removed as an (additional) mechanism for the disappearance of this population in nearby clusters. They explain the large number of edge-on systems as possibly gravitationally lensed small arcs and arclets.

Fukugita et al. (1995) studied galaxies in cluster CL0939+472 with ground based data, using simple photometric parameters. They classify the galaxies morphologically into early and late types, and comparison with HST data shows that they have a success rate of about 70%.

Recently, Belloni (1995), Belloni et al. (1995) and Belloni & Röser (1996) obtained low resolution spectra in CL0939+472 of all objects brighter than R=22.5 mag in a (5 arcmin)$^2$ field. For 85% of these objects template-fitted redshifts and morphological information could be obtained. From a total of 323 such (template-fitted) spectra of objects in their field they establish 174 galaxies as cluster members. Among them they detect 35 E+A galaxies (= post-starburst galaxies) (corresponding to about 20%), which all are disk systems (spirals and irregulars) or mergers, confirming 6 E+A galaxies previously known, and finding 29 new ones. Whereas the elliptical galaxies show an obvious concentration towards the cluster centre, E+A galaxies and spirals are more spread out, clearly less concentrated towards the cluster centre than the ellipticals. Belloni et al. (1995) show that the morphology indicates that the E+A activity is associated with mergers or disk-like systems. In contrast to Dressler et al. (1994a,1994b) and Dressler (1995), they conclude that the Butcher-Oemler effect in CL0939+472 is mainly caused by galaxies undergoing star formation, and that the effect of normal spiral galaxies is only secondary.

## 3. ROSAT Observation

CL0939+472 was observed in November 1991 with the ROSAT/PSPC with an exposure time of 14.35 ksec in a pointed observation. Figure 1 shows an image in the hard band (0.5 -2.0 keV). The side length is about 8 arcmin. We use a Gaussian filter with $\sigma = 15$ arcsec to smooth the data.

The X-ray emission of CL0903+472 (Fig. 1) clearly shows substructure, the hot gas of the cluster is not very centrally concentrated. There are three maxima visible in the central part of Fig. 1: M1: ($\alpha = 09^h42^m59^s, \delta = 46°59'18''$ (J2000)), M2: ($09^h43^m05^s, 46°59'51''$), and M3: ($09^h42^m58^s, 47°01'03''$). But when subdividing the observation into different time intervals, one finds that the two main maxima M1 and M2 are only statistical fluctuations and not significant. The only structures that seems to be robust are the maximum in the northwest M3 and the main structure of the cluster which is more or less a plateau in southeastern - northwestern direction. The emission is quite extended.

While the maximum M3 looks slightly elongated, there are three other sources close to the cluster that are point sources and probably not associated with the cluster. The first of these point sources is visible at the left side of Fig. 1, the other two are outside the boundaries of Fig. 1. The positions of these point sources are: Point source P1: position: ($09^h43^m18^s, 47°00'50''$), count rate: $f_{P1} = 5.9 \cdot 10^{-3}$ counts/s, distance from cluster centre $r_{P1} = 3.0$ arcmin; P2: ($09^h42^m34^s, 47°02'28''$) $f_{P2} = 9.7 \cdot 10^{-3}$ counts/s, $r_{P2} = 5.8$ arcmin; P3: ($09^h42^m58^s, 47°06'18''$), $f_{P3} = 6.2 \cdot 10^{-3}$ counts/s, $r_{P3} = 6.2$ arcmin. All coordinates are for equinox J2000, the count rates are calculated in the ROSAT band (0.1-2.4 keV) within a radius of 0.75 arcmin around the source which contains more than 99% of the point spread function. A source at $\alpha = 09^h42^m05^s \delta = 47°22'04''$ (J2000) coincides with the position of the G5 star AG+47 835. This X-ray source has a count rate within a radius of 1.75 arcminutes of $1.1 \cdot 10^{-2}$ counts/s. From the offset of the X-ray and the optical position we infer that the positional uncertainty of the PSPC pointing is less than 8 arcseconds.

We can trace the X-ray emission of CL0939+472 out to a radius of 3.3 arcmin (1.3 $h_{50}^{-1}$ Mpc) (see Fig. 2). Within this radius the cluster has a count rate of $0.076 \pm 0.003$ counts/s (excluding the count rate of P1). This corresponds to an X-ray luminosity in the ROSAT band (0.1-2.4 keV) of $6.7 \pm 0.3 \cdot 10^{44}$ erg/s assuming a cluster temperature of 2.9 keV, a metallicity of 0.35 (in solar units), and the galactic hydrogen column density of $n_H = 0.127 \cdot 10^{21}$ cm$^{-2}$ (Dickey & Lockman 1990). For an assumed temperature of 10 keV, the luminosity in the ROSAT band would be only higher by 1%, i.e. the luminosity depends only very weakly on the assumed cluster temperature. The bolometric luminosity for an assumed temperature of 2.9 [10] keV is $(7.5 \pm 0.3) \cdot 10^{44}$ [$(8.0 \pm 0.3) \cdot 10^{44}$] erg/s. The peak brightness for the absolute maximum M1 in Fig. 1 is $7.3 \cdot 10^{-3}$ counts/s/arcmin$^2$ (for a Gaussian smoothing with $\sigma = 15$ arcsec). The central cooling time of the cluster is $3 \cdot 10^{10}$ years. This large number indicates that only a small amount of the thermal energy has been lost by cooling.

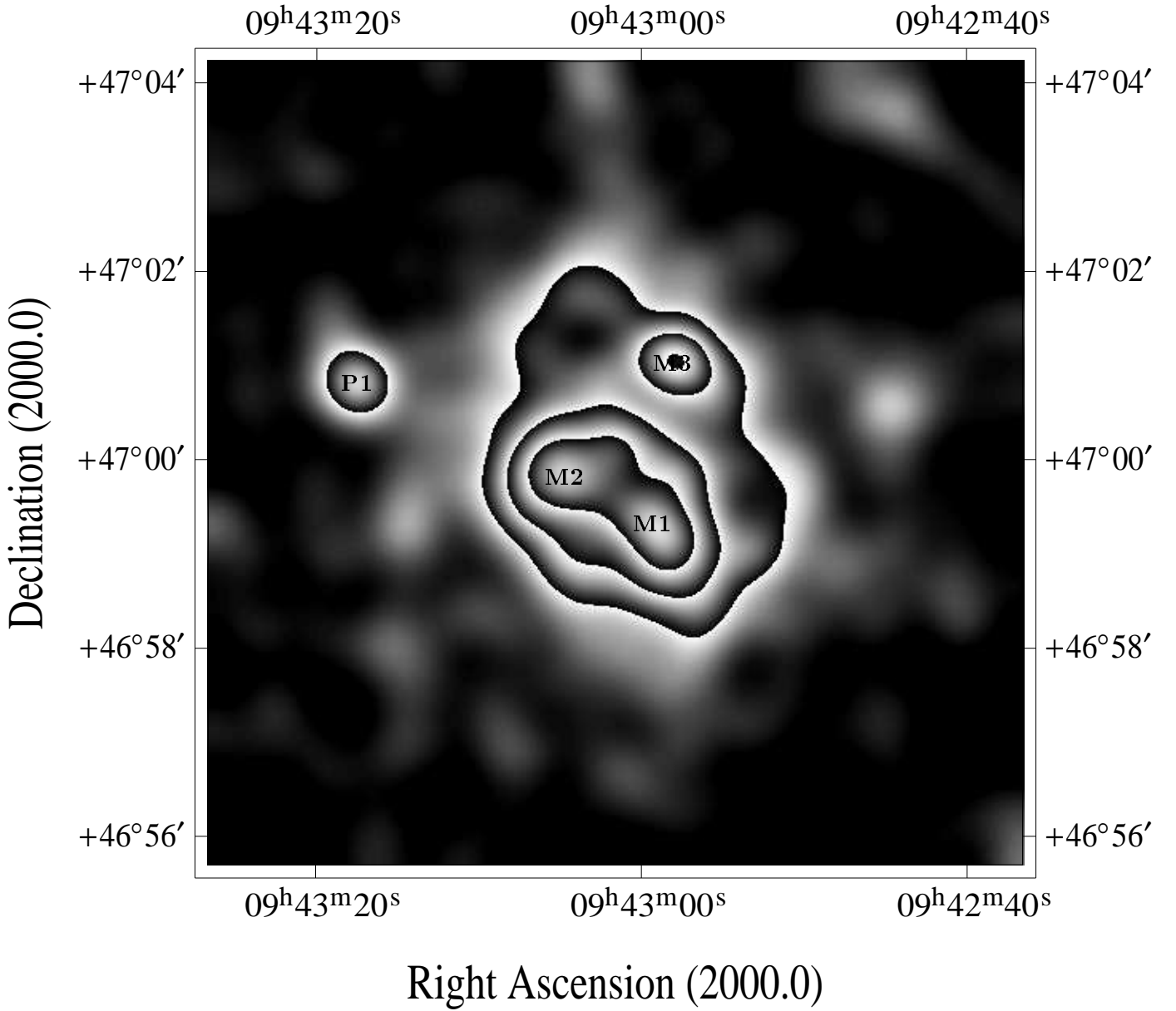

**Fig. 1.** ROSAT/PSPC image of the cluster CL0939+472 in the ROSAT hard band (0.5-2.0 keV). The data are smoothed with a Gaussian filter of $\sigma = 15$ arcsec. Three maxima (M1, M2, M3) and a probable foreground point source (P1) are marked.

The quasar behind the cluster (Dressler et al. 1993) could contribute to the X-ray luminosity. But from the relation between optical and X-ray luminosities of quasars (La Franca et al. 1995) only a contribution of about 5% of the total cluster luminosity is expected from the quasar. As there is no enhanced emission seen at the position of the quasar (see Fig. 4), it is unlikely that the quasar contribution to the total X-ray luminosity is higher than this expected value.

## 4. X-ray Profile and Mass Determination

The structure of the X-ray image is influenced by statistical fluctuations so that no detailed morphological analysis is possible. An attempt to extract some quantitative information was made by fitting ellipses to different isophote levels of the cluster (Bender & Möllenhoff 1987). We find for the outer part (up to a surface brightness of $3.5 \cdot 10^{-3}$ counts/s/arcmin$^2$) small eccentricities $\epsilon = (a-b)/a$ between 0.0 and 0.2. The plateau (brightness levels larger than $3.5 \cdot 10^{-3}$ counts/s/arcmin$^2$) shows a strong increase of the eccentricity from 0.2 up to 0.75 with a position angle of 52° (counterclockwise, north=0).

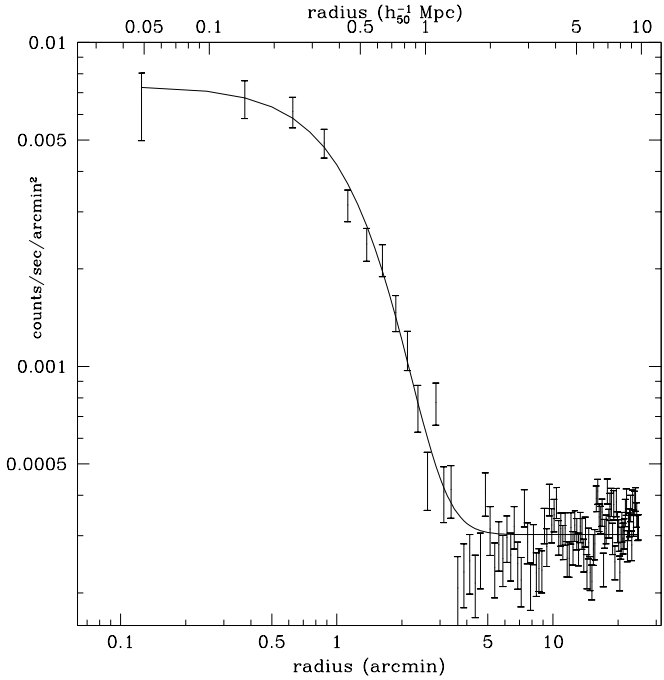

**Fig. 2.** Radial profile of the X-ray emission of cluster CL0939+472. The centre of the cluster is assumed at the position $\alpha = 09^h 43^m 02^s$ $\delta = 46°59'41''$ (J2000). The solid line is a fit with a $\beta$-model. Because of the flat extended maximum we find a huge core radius of $r_c = 1.1 h_{50}^{-1}$ Mpc.

Although some parts of the cluster are clearly not spherically symmetric, we tried to fit the surface brightness profile with a $\beta$-model (following Cavaliere & Fusco-Femiano 1976; Jones & Forman 1984):

$$\Sigma(r) = \Sigma_0 \left(1 + \left(\frac{r}{r_c}\right)^2\right)^{-3\beta+1/2}, \quad (1)$$

where $\Sigma_0$ is the central surface brightness, $r_c$ is the core radius, and $\beta$ is the slope parameter. The point sources mentioned above are excluded for the fit. The plateau-like structure yields a huge core radius and as a consequence the slope outside the core is very steep, i.e. the value of $\beta$ is very large. For the hard ROSAT band (0.5-2.0 keV) we find $\Sigma_0 = 7.0 \cdot 10^{-3}$ counts/s/arcmin$^2$, $r_c = 1.1 h_{50}^{-1}$Mpc, and $\beta$=1.9 when using as centre of the cluster $\alpha = 09^h 43^m 02^s$ $\delta = 46°59'41''$ (J2000) and a binning of 15 arcseconds for the profile (see Fig. 2). This position appears as the maximum after a coarse smoothing of the data ($\sigma$ = 30 arcsec). It is located between M1 and M2. If the maximum M1 is chosen as centre, the resulting parameters differ at most by 10%. Also, using a different binning or using the ROSAT broad band does not change the parameters considerably. Given these unusual numbers for the core radius $r_c$ and the exponent $\beta$, it is probable that there is some substructure hidden in the data which could not be resolved due to the limited resolution of the ROSAT/PSPC or the low number of source counts.

Under the assumption that the cluster is spherically symmetric, the parameters of this $\beta$-fit can be used to make a deprojection of the 2D image to get the three dimensional density distribution. (Although the cluster apparently is not spherically symmetric, such an assumption may help to get a rough estimate of the mass distribution.) With the additional assumption of hydrostatic equilibrium, the integrated mass can be calculated from the equation

$$M(r) = \frac{kr}{\mu m_p G} T \left(\frac{d \ln \rho}{d \ln r} + \frac{d \ln T}{d \ln r}\right), \quad (2)$$

where $\rho$ and $T$ are the density and the temperature of the intra-cluster gas, and $r$, $k$, $\mu$, $m_p$, and $G$ are the radius, the Boltzmann constant, the molecular weight, the proton mass, and the gravitational constant, respectively. With substructure, however, one has to be cautious in deriving the cluster mass from the spherically symmetric density distribution. Substructure yields a shallower density profile (large $r_c$) e.g. a smaller density gradient and leads therefore to an underestimation of the total mass (Schindler 1996). The mass determined for CL0939+472 by Eq. 2 is therefore only a lower limit.

The temperature of the intra-cluster gas needed to determine the mass from Eq. 2 is derived by spectral fitting. For the spectral analysis we use the photons within a radius of 2.3 arcminutes around the central position ($09^h 43^m 02^s$, $46°59'41''$) in the energy range between 0.2 – 2.4 keV. The spectrum is rebinned at a minimum signal-to-noise ratio of five in each of the spectral bins. We fit the spectrum with a Raymond-Smith model (Raymond & Smith 1977). As the spectrum contains only about 1000 source counts, the number of fit parameters has to be reduced to a minimum. The hydrogen column density is fixed to the galactic value $n_H = 0.127 \cdot 10^{21}$ cm$^{-2}$ (Dickey & Lockman 1990) and for the redshift z=0.407 (Dressler & Gunn 1992) is used. We find a temperature of $2.9^{+1.3}_{-0.8}$ keV ($1\sigma$ errors, reduced $\chi^2$=1.0, number of degrees of freedom = 21). The temperature is independent of the assumed metallicity (the metallicity could not be determined with this spectrum). Neither a different binning, nor a different energy range, nor different background assumptions change the result of the temperature determination significantly.

Although the result for the temperature seems to be quite robust, it is possible that ROSAT with its relatively soft energy range (0.1-2.4 keV) underestimates the temperature of hot clusters. But as we only want to give a lower limit for the mass, this does not affect our result (see Eq. 2). Due to the limited number of photons the radial dependence of the temperature cannot be determined. Therefore, we have to assume an isothermal cluster for the mass determination. The temperature gradient does not have a large effect on the mass determination for most clusters as the density gradient was found to be generally the dominant one (Schindler 1996). The only possibility how the mass could be overestimated by assuming an isothermal cluster is if the temperature was decreasing towards the centre i.e. if there was a cooling flow in the cluster. But from the shallow profile, the plateau-like structure and the large central cooling time this possibility can be clearly ruled out. Therefore using an isothermal cluster at 2.9 keV is a good assumption for an estimate of a lower limit of the mass.

The integrated (three-dimensional) mass profile $M(r)$ as estimated from Eq. 2 with the above mentioned parameters is shown in Fig. 3 as the thick solid line. The uncertainty of the integrated mass profile (indicated with the thin solid lines) is +45%/−30%, largely reflecting the temperature errors (the uncertainty in the density gradient is negligible compared to that of the temperature).

converges with the solid line (total mass) at large radii (outside the range of the diagram). The uncertainty in the surface mass density is the same as the one for the total mass mentioned above.

In Table 1 the numerical values of the mass determinations are given for radius values of $0.5 h_{50}^{-1}$ Mpc and $1 h_{50}^{-1}$ Mpc, respectively (uncertainties for the total mass and the integrated surface mass density are, as stated above, +45%/−30%).

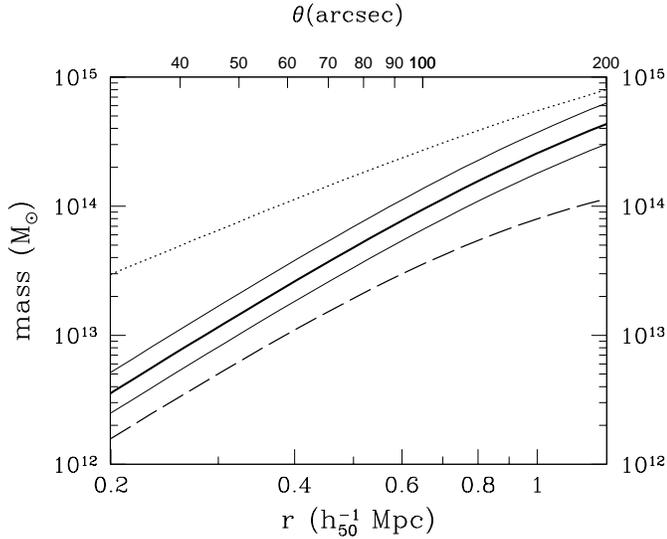

|  | mass within $0.5 h_{50}^{-1}$ Mpc | mass within $1.0 h_{50}^{-1}$ Mpc |
|---|---|---|
| gas mass | $0.19 \times 10^{14}$ M$_\odot$ | $0.84 \times 10^{14}$ M$_\odot$ |
| total mass M(r) | $0.48 \times 10^{14}$ M$_\odot$ | $2.6 \times 10^{14}$ M$_\odot$ |
| integrated surface mass density $\int \Sigma(\theta)d\theta$ | $1.7 \times 10^{14}$ M$_\odot$ | $5.5 \times 10^{14}$ M$_\odot$ |

**Table 1.** Numerical values for the mass determination in cluster CL0939+472 for two specific radii; for the total mass and the integrated surface mass density these values are lower limits.

**Fig. 3.** Mass estimates of cluster CL0939+472 as function of radius: The dashed line shows the integrated mass of the X-ray gas M$_{gas}$(r) assuming a spherically symmetric mass distribution. The thick solid line marks the lower limit of the total mass M(r); the thin solid lines indicate the uncertainties of +45%/-30%. For comparison with mass determinations by the gravitational lens effect, the dotted line depicts the integrated surface mass density M($\theta$) = $\int \Sigma(\theta)d\theta$.

As the emissivity of the gas in the ROSAT energy band is almost independent of the temperature (within the temperature range of 2-10 keV it changes only by 6%), we can derive the gas density distribution without the uncertainty of the temperature estimate. The only uncertainty are local inhomogeneities of the gas. This uncertainty is hard to quantify, but is certainly smaller than the uncertainty in temperature. The integrated gas mass is shown as the dashed line in Fig. 3. The gas mass fraction tends to decrease slightly outward.

For a comparison with mass estimates determined from the weak gravitational lensing effect in the cluster (cf. Seitz et al. 1995), we also calculate the total mass of the cluster, as seen inside a certain angle, basically integrating a spherically symmetric three dimensional mass distribution in cylindrical shells with cylinder axis parallel to the line of sight (dotted line), or integrating the cluster surface mass density $\Sigma(\theta)$ outward (for the conversion from $\theta$ in arcseconds into $r$ in Mpc in Fig. 3 we assume $H_0$ = 50 km/s/Mpc and $q_0$ = 0.5). This dotted line

## 5. Discussion and Conclusions

There are several hints that CL0939+472 is not a standard virialised cluster: as there exists a clear correlation between galaxy number density and luminosity (Mushotzky 1984; Edge & Stewart 1991b) one would expect for this extremely rich cluster an extremely high X-ray luminosity. But with its bolometric luminosity of $8 \cdot 10^{44}$ erg/s, it is not among the most luminous clusters. For comparison, the rich, distant cluster CL00016+16 has an X-ray luminosity about 4 times higher (White et al. 1981). Furthermore, the X-ray morphology shows clearly that CL0939+472 is not relaxed: there is substructure and an extended plateau-like structure.

The temperature of 2.9 keV is relatively low. Compared with the $L_X$-$T_X$ relation (Edge & Stewart 1991a; David et al. 1993) and the $\sigma$-$T_X$ relation (Edge & Stewart 1991a; Lubin & Bahcall 1993; Bird et al. 1995) one would expect a temperature between 4 and 5 keV. The same number is expected from the velocity dispersion with a virial assumption. One possible explanation is that the temperature is underestimated by the ROSAT/PSPC data. Another explanation is a redshift dependence of the $L_X$-$T_X$ relation (see Henry et al. 1994). A third possibility is that the relation does not apply to this cluster because it is not virialised but rather in the process of merging.

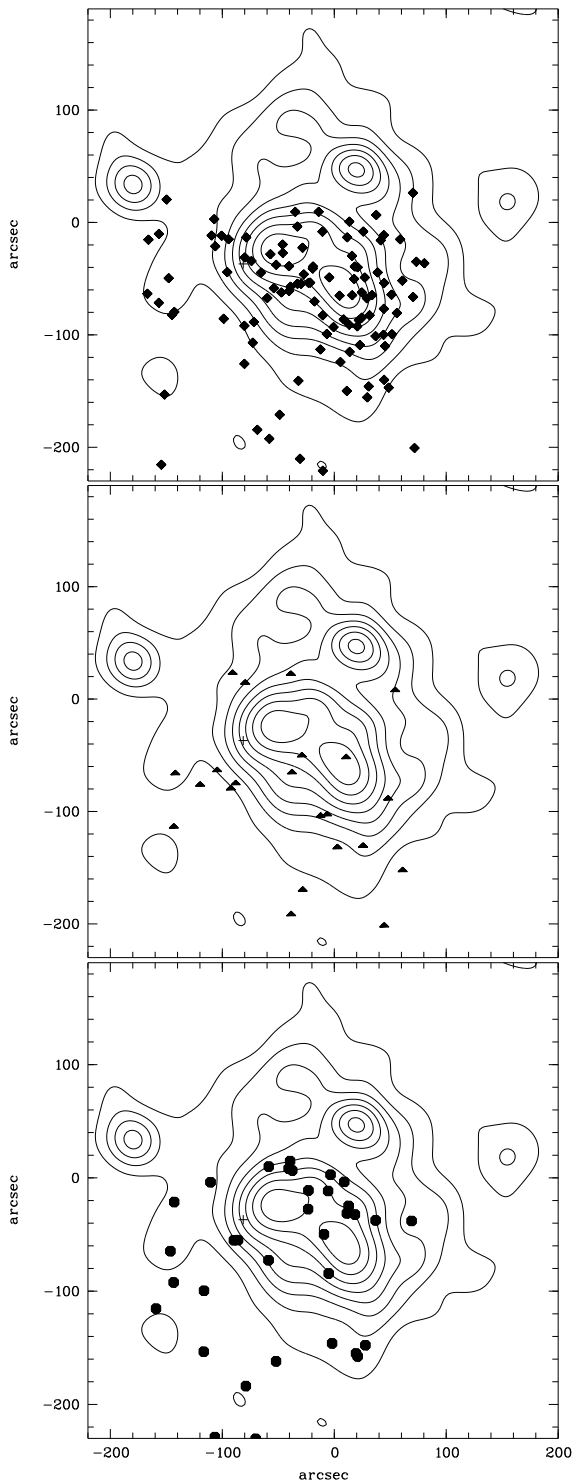

**Fig. 4.** Cluster galaxies as identified by Belloni & Röser (1996) superimposed on the X-ray contours of the ROSAT hard band (0.5-2.0 keV) of cluster CL0939+472 : top: elliptical galaxies (diamonds); centre: spiral and irregular galaxies (triangles); bottom: E+A galaxies (octagons). They identified the galaxies in a 5'x5' field which covers only part of the X-ray emission. The X-ray image is smoothed with a Gaussian with $\sigma = 15$ arcsec (North is up, East is left). The contours are linearly spaced with $\Delta$count rate = $8.1 \cdot 10^{-4}$ counts/s/arcmin$^2$ the highest contour line corresponding to $6.5 \cdot 10^{-3}$ counts/s/arcmin$^2$. The position of the z=2.05 quasar (Dressler et al. 1993) is marked with a plus sign just left of the centre.

Another hint that the cluster is dynamically young is the morphological mix and the spatial distribution of galaxies. There is a very high fraction of spiral/irregular and of E+A galaxies in CL0939+472. The spatial distribution of the different galaxy types (cf. Belloni & Röser 1996) is shown in Fig. 4. The ellipticals are concentrated around the elongated X-ray maximum. It is expected from the the density-morphology relation (Dressler 1980) and the radius-morphology relation (Whitmore et al. 1993) that the ellipticals trace the cluster centre best. But seemingly the main concentration of ellipticals is slightly west of M1, as well as the cluster centre given by Dressler & Gunn (1992). In contrast to the ellipticals, the E+A galaxies are concentrated north of the X-ray maximum indicating a possible merger activity in the recent past. The analysis of the galaxy types in the northern subcluster M3 (Belloni, in preparation) will provide more information on this part of the cluster.

The typical mass range of clusters within the virial radius is $5 - 50 \cdot 10^{14} M_\odot$ (Böhringer 1996). For a comparison we calculate the virial radius of CL0939+472 with the same spherical collapse approximation (Gunn & Gott 1972; White et al. 1993) as in Böhringer (1996). We find a radius of about 3 Mpc. An extrapolation of the integrated total mass to this radius yields $1.5 \cdot 10^{15} M_\odot$ which is well within the typical mass range. A comparison of the gas mass with the lower limit of the total mass gives an upper limit for the gas mass fraction. In CL0939+472 we find a gas mass fraction between 25% and 50%, which is higher than for most other clusters (Böhringer 1996). Compared with the relation of bolometric X-ray luminosity and gas mass within 0.5 Mpc (Edge & Stewart 1991a), the gas mass is slightly on the small mass side.

It is fair to say that this cluster seems to be "abnormal" in a couple of respects, as mentioned at various sections in this paper. However, all these unusual properties can be easily explained in one scenario: if the cluster is in fact undergoing a merger event, i.e. two clusters of roughly similar mass are colliding and forming a very rich new cluster. This could explain the high spiral fraction in the central region, and the merger event would initiate activity processes and could thus explain the high fraction of E+A galaxies. The clear substructure in the X-ray morphology is also most easily explained as reflecting the unrelaxed gas in or shortly after an collision process. The sharp drop off of the plateau could be an indication for a shock wave, which is expected from numerical simulations (Schindler & Müller 1993) after the collision of subclusters. Furthermore the asymmetric redshift distribution of the galaxies as seen in Belloni et al. (1995) supports this view. With a high-resolution pointed observation of this cluster by the ROSAT/HRI instrument, such a merger scenario can possibly be confirmed.

*Acknowledgements.* It is a pleasure to thank Paola Belloni, Makoto Hattori, Hans Böhringer, and Doris Neumann for helpful discussions. We also thank Paola Belloni for making available to us the galaxy data in electronic form. The generous hospitality of the Astronomical Institute of the University of Basel is greatly appreciated. S.S. acknowledges financial support by the Verbundforschung.